\documentclass[twocolumn,showpacs,prl]{revtex4}
\usepackage{graphicx}
\newcommand{\ket}[1]{|{#1}\rangle}			
\newcommand{\bra}[1]{\langle{#1}|}			
\newcommand{\ketbra}[3]{\ket{#1}_{#3}\bra{#2}}		
\newcommand{\braket}[1]{\langle {#1}\rangle}

\begin{document}
\title{Unconditionally Secure Key Distribution Based on Two Nonorthogonal 
States}

\author{Kiyoshi Tamaki}
\author{Masato Koashi}
\author{Nobuyuki Imoto}
\affiliation{CREST Research Team for Interacting Carrier Electronics, School
of Advanced Sciences,\\ The Graduate University for  Advanced Studies
(SOKENDAI), Hayama, Kanagawa, 240-0193, Japan}


\begin{abstract} 
We prove the unconditional security of the Bennett 1992 protocol, by
using a reduction to an entanglement distillation protocol initiated by a
local filtering process. The bit errors and the phase
errors are correlated after the filtering, and we can bound the amount 
of phase errors from the observed bit errors by an estimation method 
involving nonorthogonal measurements. The angle between the two states
shows a trade-off between accuracy of the estimation and robustness to
noises.

\end{abstract}

\maketitle

Quantum key distribution (QKD) provides a way to share a secret key
between two parties (Alice and Bob) with very small leak of 
information to an eavesdropper (Eve). One of the simplest of such protocols
is called B92 \cite{B92}, which is based on the transmission of only two 
nonorthogonal states. For a qubit channel between Alice and Bob,
this protocol proceeds as follows. Alice randomly chooses a bit value $j$,
and prepare a qubit in state  
$\ket{\varphi_j}\equiv\beta\ket{0_x}
+(-1)^j\alpha\ket{1_x}$, where $0<\alpha< 1/\sqrt{2}$, 
$\beta\equiv\sqrt{1-\alpha^2}$, and 
$\{\ket{0_x},\ket{1_x}\}$ is a basis ($X$-basis) of the qubit.
She sends the qubit through the channel to Bob, who 
performs a measurement ${\cal M}_{\rm B92}$ 
with three outcomes $j'=0,1, {\rm ``null"}$.
The measurement ${\cal M}_{\rm B92}$ is defined by the POVM
$F_{0}=\ket{\overline{\varphi}_1}\bra{\overline{\varphi}_1}/2$,
$F_{1}=\ket{\overline{\varphi}_0}\bra{\overline{\varphi}_0}/2$,
and $F_{\rm null}=1-F_{0}-F_{1}$, where 
$\ket{\overline{\varphi}_j}\equiv\alpha\ket{0_x}
-(-1)^j\beta\ket{1_x}$
is the state orthogonal to $\ket{\varphi_j}$. When the outcome is
$j'={\rm null}$, Bob announces that to Alice and they discard the event.
Otherwise, they take notes of their bit values $j$ and $j'$, which should 
coincide in the absence of channel noises and Eve's intervention.
Repeating this procedure many times, each of Alice and Bob obtains 
a sequence of bits. Then they converts the sequences into 
a shared secret key through public discussions.

Although the QKD protocols themselves
are simple, proving the unconditional security is quite hard, 
since Eve may make a very complicated attack such as 
interacting all of the transmitted qubits jointly
to a big probe system. This task has been accomplished
\cite{m98} for the BB84 protocol \cite{BB84},
which involves four states forming two conjugate bases.
Subsequent proofs \cite{others,lc98,sp00} have provided us more than a
basic claim of security, including a beautiful
interplay 
\cite{lc98,sp00} between QKD and other important protocols
 in quantum information, such as the entanglement distillation protocol
 (EDP) \cite{EPP}
and the Calderbank-Shor-Steane (CSS) 
quantum error correcting codes \cite{CSS}.
It is natural to ask about the unconditional security of the B92 protocol,
which is conceptually the simplest of the QKD protocols. In contrast 
to BB84, it involves a free parameter $\alpha$ representing the 
nonorthogonality. The analyses of the B92 protocol is hence expected 
to give us an idea about how the nonorthogonality is related to the 
ability to convey secret information. Since the security proofs of 
BB84 rely on the symmetry of the protocol which is not shared in 
B92, it is not a trivial task to modify it for B92, except for 
the limiting case of $|\braket{\varphi_0|\varphi_1}|^2=1/2$ \cite{qc02}.

In this Letter, we give a proof of the unconditional security of 
the B92 protocol for qubit channels, applicable to any amount of 
nonorthogonality $\alpha$. We show that the B92 protocol is 
related to an EDP initiated by a local filtering \cite{Gisin96}.
We also develop a method to estimate an error rate by
measuring randomly chosen samples on a different basis, which 
plays an important role in the proof.
 
We first introduce a protocol involving EDP, which is 
then shown to be reduced to the B92 protocol.
We assume that Alice initially prepares a pair of qubits AB
in the 
state
$\ket{\Psi}_{\rm{AB}}=\left(\ket{0_{z}}_{\rm{A}}\ket{\varphi_0}_{\rm{B}}
+\ket{1_{z}}_{\rm{A}}\ket{\varphi_1}_{\rm{B}}\right)/\sqrt{2}$, 
which is nonmaximally entangled. Here $Z$-basis $\{\ket{0_z},\ket{1_z}\}$
of a qubit is related to the $X$-basis by 
$\ket{j_z}=[\ket{0_x}+(-1)^j\ket{1_x}]/\sqrt{2}$.
Alice sends Bob the qubit B through a 
quantum channel. 
Suppose that Bob performs
a ``local filtering operation'' on qubit B, described by 
the Hermitian operator $F_{\rm{fil}}\equiv \alpha\ket{0_x}_{\rm
B}\bra{0_x}+
\beta\ket{1_x}_{\rm B}\bra{1_x}$. When the state of AB was $\rho$, the 
qubit B passes the filtering with probability 
$p={\rm Tr}[\rho({\bf 1}_{\rm A}\otimes F_{\rm{fil}})^2]$,
resulting in the filtered state $[({\bf 1}_{\rm A}\otimes F_{\rm{fil}})
\rho({\bf 1}_{\rm A}\otimes F_{\rm{fil}})]/p$. 
When the channel is noiseless and Eve does nothing, this process is 
just the Procrustean method mentioned in \cite{bbps96}: the filtered state
 should be the maximally entangled state (EPR state)
$\ket{\Phi^+}=(\ket{0_{x}}_{\rm{A}}\ket{0_{x}}_{\rm{B}}
+\ket{1_{x}}_{\rm{A}}\ket{1_{x}}_{\rm{B}})/\sqrt{2}$,
since the initial state is 
also written as  
$\ket{\Psi}_{\rm{AB}}=\beta\ket{0_{x}}_{\rm{A}}\ket{0_{x}}_{\rm{B}}
+\alpha\ket{1_{x}}_{\rm{A}}\ket{1_{x}}_{\rm{B}}$.
When noises are present, the filtered state may include 
a bit error, represented by the subspace spanned by
$\{\ket{0_{z}}_{\rm{A}}\ket{1_{z}}_{\rm{B}},\ket{1_{z}}_{\rm{A}}\ket{0_{z}}_{\rm{B}}\}$,
and a phase error, represented by the subspace spanned by
$\{\ket{0_{x}}_{\rm{A}}\ket{1_{x}}_{\rm{B}},\ket{1_{x}}_{\rm{A}}\ket{0_{x}}_{\rm{B}}\}$.
In parallel to the protocols for BB84 \cite{lc98, sp00},
we can consider the following 
 protocol that will work under the presence of
noises. 

 {\em Protocol 1:} (1) Alice creates $2N$ pairs in the
state $\ket{\Psi}_{\rm{AB}}^{\otimes 2N}$, and she sends the second half
of each pair to Bob over a quantum channel.  (2) By public discussion, 
Alice and Bob randomly permute the position of $2N$ pairs of qubits.
(3) For the first $N$ pairs (check pairs), Alice measures her halves on $Z$-basis,
and Bob performs measurement ${\cal M}_{\rm B92}$ on his halves.
By public discussion, they determine the number $n_{\rm err}$ of errors 
in which Alice found $\ket{0_z}$ and Bob's outcome was $1$, or 
Alice found $\ket{1_z}$ with Bob's outcome $0$.
(4) For the second $N$ pairs (data pairs), Bob performs the filtering
$F_{\rm{fil}}$
 on each of his qubits, and 
 announces the total number $n_{\rm fil}$ and 
the positions of the qubits that have passed the filtering.
(5) From $n_{\rm err}$ and $n_{\rm fil}$, they estimate an upper
bound for the number of bit errors $n_{\rm bit}$,
 and an upper bound for the number of phase errors $n_{\rm ph}$,
in the $n_{\rm fil}$ pairs.
If these bounds are too
large, they abort the protocol.
(6) They run an EDP that can produce $n_{\rm key}$ nearly perfect 
EPR pairs if the estimation is correct.
(7) Alice and Bob each measures the EPR pairs in $Z$-basis to
obtain a shared secret key.

For the same reason as in the proofs of BB84 \cite{lc98, sp00},
if the estimation in step (5) is correct except for a 
probability that becomes exponentially small as $N$ increases,
the final shared key is essentially secure. Intuitively, this 
comes from the fact that Eve has no clue on the outcomes of 
a measurement performed on an EPR pair, since it is in a pure 
state by definition. We will soon show how to estimate the 
upper bounds for the errors in step (5). Before that, we will
show that {\em Protocol 1} can be reduced to the B92 protocol.

According to the discussion by Shor and Preskill \cite{sp00},
we can use a one-way EDP based on CSS codes in step (6). Then,
they have further shown that
the whole extraction process of the $n_{\rm key}$-bit final secret key 
from the noisy $n_{\rm fil}$ pairs in steps (6) and (7) can be
equivalently accomplished by $Z$-basis measurements directly 
performed on Alice's and Bob's qubits of the $n_{\rm fil}$ noisy pairs, 
followed by a public discussion. Hence, without affecting the security,
we can assume that Alice performs $Z$-basis measurements immediately 
after she has prepared the state $\ket{\Psi}_{\rm{AB}}$, and that
Bob performs $Z$-basis measurements immediately 
after he has performed the filtering. {\em Protocol 1} is thus 
reduced to a prepare-and-measure protocol.
Now, note the following 
relation for $j'=0,1$, which is easily confirmed:
\begin{equation}
F_{\rm{fil}}\ketbra{j'_z}{j'_z}{\rm{B}}F_{\rm{fil}}
=F_{j'}.
\label{equivMfil}
\end{equation}
This implies that the filtering followed by the $Z$-basis measurement 
is, as a whole, equivalent to the measurement ${\cal M}_{\rm B92}$. 
Hence in the reduced protocol Alice simply sends $\ket{\varphi_0}$ and 
$\ket{\varphi_1}$ randomly, and Bob performs ${\cal M}_{\rm B92}$ on 
all of the received qubits, which completes the reduction to B92.

The estimation in step (5) can be done as follows. The number of 
bit errors $n_{\rm bit}$ could be determined if Alice and Bob
exchange their measurement results in $Z$-basis. But this is 
the same process as the one performed on the first $N$ pairs 
to obtain $n_{\rm err}$, due to the relation (\ref{equivMfil}).
Thanks to the random permutation in step (2), the check pairs 
are regarded as a classical random sample from the $2N$ pairs.
Then, from a classical probability estimate, we may assume 
\begin{equation}
|n_{\rm bit}-n_{\rm err}|\le N\epsilon_1.
\label{ineq1}
\end{equation}
For any strategy by Eve, the probability of violating this inequality
is asymptotically less than $\exp(-N \epsilon_1^{2})$.
 
The estimation of the phase errors is far more complicated.
To do this, we derive several inequalities by assuming
gedanken measurements that are not really done in the {\em Protocol 1}.
The number of phase errors $n_{\rm ph}$ could be determined if 
Alice and Bob measure the $n_{\rm fil}$ pairs in $X$-basis 
just after step (4). Since the filtering operator $F_{\rm fil}$ is 
also diagonal in $X$-basis, 
$n_{\rm fil}$ and $n_{\rm ph}$ could also be determined 
by another measurement scheme, in which Alice and Bob
perform $X$-basis measurements first, and then Bob applies the 
filtering $F_{\rm fil}$. Note that this filtering can be
 done classically by Bernoulli trials since the outcomes of 
the $X$-basis measurements are available.
This new scheme also produces 
the numbers $n_{ij} (i,j=0,1)$ of pairs found in 
state $\ket{i_x}_{\rm{A}}\ket{j_x}_{\rm{B}}$. 
Since $n_{ij}$ and $n_{\rm fil}$ ($n_{\rm ph}$)
 are related by Bernoulli trials,
we have 
\begin{eqnarray}
|\alpha^2(n_{00}+n_{10})+\beta^2(n_{01}+n_{11})-n_{\rm fil}|
 \le N\epsilon_2 
\label{ineq2}
\\
|\alpha^2n_{10}+\beta^2n_{01}-n_{\rm ph}|
 \le N\epsilon_3,
\label{ineq3}
\end{eqnarray}
which are violated with probability asymptotically less than 
$\exp(-2N\epsilon_{2}^2)$ and $\exp(-2N\epsilon_{3}^2)$, respectively. 

Next, recall the fact that neither the noisy channel nor Eve can 
touch the qubits held by Alice. This implies that the marginal 
state of Alice's data qubits before the measurements
should be $\rho_{\rm A}^{\otimes N}$,
where $\rho_{\rm A}\equiv {\rm
Tr_{B}}(\ketbra{\Psi}{\Psi}{\rm{AB}})=
\beta^2\ket{0_{x}}_{\rm{A}}\bra{0_{x}}
+\alpha^2\ket{1_{x}}_{\rm{A}}\bra{1_{x}}$.
We can thus regard $n_{10}+n_{11}$ as a result of a Bernoulli trial,
obtaining
\begin{equation}
|\alpha^2N-(n_{10}+n_{11})|
 \le N\epsilon_4
\label{ineq4}
\end{equation}
with probability of violation asymptotically less than 
$\exp(-2N\epsilon_{4}^2)$.

Let us switch to the measurement on the check pairs (the first
$N$ pairs).
The element of POVM corresponding to the error in step (3)
is given by $\Pi_{\rm{err}}=(\ketbra{\Gamma_{11}}{\Gamma_{11}}{}
+\ketbra{\Gamma_{01}}{\Gamma_{01}}{})/2$, where
$\ket{\Gamma_{11}}\equiv\alpha\ket{0_x}_{\rm{A}}\ket{0_x}_{\rm{B}}
-\beta\ket{1_x}_{\rm{A}}\ket{1_x}_{\rm{B}}$ and 
$\ket{\Gamma_{01}}\equiv\beta\ket{0_x}_{\rm{A}}\ket{1_x}_{\rm{B}}
-\alpha\ket{1_x}_{\rm{A}}\ket{0_x}_{\rm{B}}$.
This is readily derived from the relation
$\ket{j_z}_{\rm A}\ket{\overline{\varphi}_j}_{\rm B}
=\ket{\Gamma_{11}}-(-1)^j\ket{\Gamma_{01}}$. Let us add two more
states, $\ket{\Gamma_{00}}\equiv\beta\ket{0_x}_{\rm{A}}\ket{0_x}_{\rm{B}}
+\alpha\ket{1_x}_{\rm{A}}\ket{1_x}_{\rm{B}}$ and
$\ket{\Gamma_{10}}\equiv\alpha\ket{0_x}_{\rm{A}}\ket{1_x}_{\rm{B}}
+\beta\ket{1_x}_{\rm{A}}\ket{0_x}_{\rm{B}}$, to form a basis.
While $n_{\rm err}$ is determined from local measurements in step 3,
the same outcome could be obtained by performing globally the 
complete measurement 
on basis $\{\ket{\Gamma_{ij}}\}$, followed by a Bernoulli trial
with probability $1/2$. Let $m_{ij}$ be the number of pairs found in
$\ket{\Gamma_{ij}}$. Then we have 
\begin{equation}
|(m_{11}+m_{01})/2-n_{\rm err}|
 \le N\epsilon_5,
\label{ineq5}
\end{equation}
which is violated with probability asymptotically less than 
$\exp(-2N\epsilon_{5}^2)$.

Since $\{\ket{\Gamma_{01}},\ket{\Gamma_{10}}\}$
and $\{\ket{0_x}_{\rm{A}}\ket{1_x}_{\rm{B}},
\ket{1_x}_{\rm{A}}\ket{0_x}_{\rm{B}} \}$ span the same
subspace, we can relate $m_{10}+m_{01}$ and $n_{10}+n_{01}$
by the classical probability estimate as in Eq.~(\ref{ineq1}):
\begin{equation}
|(m_{10}+m_{01})-(n_{10}+n_{01})|
 \le N\epsilon_6,
\label{ineq6}
\end{equation}
which is violated with probability asymptotically less than 
$\exp(-N \epsilon_6^{2})$.
We would like further to relate $m_{01}/(m_{01}+m_{10})$ to
$n_{01}/(n_{01}+n_{10})$, but we can no longer apply classical arguments
 here, since $\ket{\Gamma_{01}}$
and $\ket{0_x}_{\rm{A}}\ket{1_x}_{\rm{B}}$ are nonorthogonal. We will thus
 extend the classical probability estimate to the quantum case in the 
 following.

The problem to be considered is as follows.  $M=M_0+M_1$ qubits are prepared in a state,
and the position of qubits  are then randomly permuted. Then, each of the
first $M_0$ qubits is measured on an orthogonal  basis
$\{|0,0\rangle,|0,1\rangle\}$, and the rest of 
$M_1$ qubits are measured on another basis $\{|1,0\rangle,|1,1\rangle\}$.
What we ask is the bound for the probability $p(\delta_0,\delta_1)$, with
which $M_0\delta_0$ qubits are found to be in $|0,1\rangle$ and
$M_1\delta_1$ qubits are found to be in $|1,1\rangle$. Let $\rho$ be the
state after the permutation, and 
$|\chi\rangle\equiv\bigotimes_{b,j}|b,j\rangle^{\otimes n_{b,j}}$, where 
$n_{b,1}=M_b\delta_b$ and $n_{b,0}=M_b(1-\delta_b)$. Then, the
probability is given by
\begin{equation} p(\delta_0,\delta_1)=
\langle \chi|\rho|\chi\rangle
\prod_{b=0,1}\frac{M_b!}{n_{b,0}!n_{b,1}!}
\label{pdd}
\end{equation}

The technique used \cite{KW-HM} for problems involving 
i.i.d. quantum sources is also useful here, although in our case 
the state $\rho$ may be highly correlated. 
 The Hilbert space for the $M$ qubits, ${\cal H}^{\otimes M}$, can be
decomposed as ${\cal H}^{\otimes M}\cong
\bigoplus_\lambda {\cal U}_\lambda \otimes {\cal V}_\lambda$ such that
any operator of form $U^{\otimes M}$  with $U\in SU(2)$ is decomposed as 
$U^{\otimes M}\cong \bigoplus_\lambda \pi_\lambda(U) \otimes {\bf 1}$,
and any unitary operator $S_p$ corresponding to permutation
$p\in S_M$ is decomposed as 
$S_p\cong \bigoplus_\lambda {\bf 1}\otimes  \tilde\pi_\lambda(p)$. Here
the maps $\pi_\lambda$ and $\tilde\pi_\lambda$ are irreducible
representations of $SU(2)$ and $S_M$, respectively. The index $\lambda$
runs over all Young diagrams with two rows and 
$M$ boxes, namely, $\lambda=(M-k,k)$ with $k=0,1,\ldots, \lfloor M/2
\rfloor$. We will thus use $k$ instead of $\lambda$ below. For later use,
we derive a convenient form of the projection $P_k$ onto ${\cal U}_k
\otimes {\cal V}_k$.  Let us parameterize the pure states of a qubit as
$|{\bf n}\rangle$, using the unit vector ${\bf n}$ in the Bloch sphere.
 Define  a state on ${\cal H}^{\otimes M}$ as 
$|k,{\bf n}\rangle\equiv |\Psi\rangle^{\otimes k}|{\bf n}\rangle^{\otimes
M-2k}$, where $|\Psi\rangle$ is the singlet state 
$(|0\rangle|1\rangle-|1\rangle|0\rangle)/\sqrt{2}$ of two qubits. The
state $|k,{\bf n}\rangle$ is contained in subspace
${\cal U}_k \otimes {\cal V}_k$. Consider the operator with unit trace
\begin{equation}
\frac{1}{4\pi M!}\sum_p \int d{\bf n} S_p|k,{\bf n}\rangle\langle k,
{\bf n}|S_p^\dagger.
\label{projection}
\end{equation} Since it commutes with any $S_p$ and any $U\in SU(2)$, it
should be equal to $(d_k^{\cal U} d_k^{\cal V})^{-1} P_k$, where
$d_k^{\cal U}\equiv {\rm dim} {\cal U}_k$  and $d_k^{\cal V}\equiv {\rm
dim} {\cal V}_k$.

Since $\rho$ commutes with any $S_p$, it can be decomposed as 
$\rho \cong \bigoplus_k (p_k/d_k^{\cal V}) \rho_k
\otimes {\bf 1}$, where $\sum p_k=1$ and ${\rm Tr}\rho_k=1$. Then, 
$\langle \chi|\rho|\chi\rangle\le \sum_k (p_k/d_k^{\cal V})
\langle \chi|P_k|\chi\rangle$. Substituting the  form of
(\ref{projection}) to $P_k$, we have
\begin{equation}
\langle \chi|\rho|\chi\rangle \le 
\max_{k, {\bf n}}
\frac{d_k^{\cal U}}{ M!}
\sum_{p}  |\langle \chi|S_p|k,{\bf n}\rangle|^2
\end{equation}

Recall that $|\chi\rangle$ takes the form of 
$|\chi\rangle=\bigotimes_{\nu}|\nu\rangle^{\otimes n_{\nu}}$, where $\nu$
represents the double index
$(b, j)$. Then, $|\langle \chi|S_p|k,{\bf n}\rangle|^2$ becomes the
product of $(S_{\nu\nu'})^{s_{\nu\nu'}}$ and 
$(T_{\nu})^{t_{\nu}}$, where $S_{\nu\nu'}\equiv |\langle \nu |\langle
\nu'||\Psi\rangle|^2$ and 
$T_{\nu}\equiv |\langle \nu |{\bf n}\rangle|^2$. The numbers
$s_{\nu\nu'}$ and $t_{\nu}$ depend  on the permutation $p$. Let
$\mu(\{s_{\nu\nu'}\},\{t_{\nu}\})$ be the number of different
permutations that give  the same values of $\{s_{\nu\nu'}\},\{t_{\nu}\}$.
Explicitly, this degeneracy factor is given by
\begin{equation}
\mu=\left(\prod_\nu n_\nu!\right)
\frac{k!}{\prod_{\nu,\nu'}s_{\nu\nu'}!}
\frac{(M-2k)!}{\prod_{\nu}t_{\nu}!}
\label{mu}
\end{equation} Using this factor, the summation over $p$ can be replaced
by  the summation over $\{s_{\nu\nu'}\},\{t_{\nu}\}$, which  take at most
$poly(M)$ values. Since $d_k^{\cal U}=M-2k+1$ is also $poly(M)$, we obtain
\begin{equation}
\langle \chi|\rho|\chi\rangle \le poly(M)
\max_{k, {\bf n},\{s_{\nu\nu'}\},\{t_{\nu}\}}
\frac{\mu}{ M!} 
\prod_{\nu,\nu'}(S_{\nu\nu'})^{s_{\nu\nu'}}  
\prod_{\nu}(T_{\nu})^{t_{\nu}}
\label{upper}
\end{equation} Combining the Eqs.~(\ref{pdd}), (\ref{mu}), and
(\ref{upper}), and replacing the factorials by the entropy function 
$H(p_i)\equiv -\sum_i p_i\log p_i$ using the formula 
$poly(N)^{-1} \le \exp[-NH(p_i)]N!/\prod (N p_i)!  \le 1$, we can cast
the upper bound into the form 
$p(\delta_0,\delta_1)\le poly(M) \exp[- M \min R]$, where the exponent
$R$ is given by 
\begin{eqnarray} R&=&H(M_b/M)+(k/M)[D(s_{\nu\nu'}/k|S_{\nu\nu'}/4)-2]
\nonumber \\ &&+(1-2k/M)[D(t_{\nu}/(M-2k)|T_{\nu}/2)-1],
\label{R1}
\end{eqnarray} where $D$ is the relative entropy defined by 
$D(p_i|q_i)=\sum_i p_i \log_2 (p_i/q_i)$. The empirical probability
$p_{bj}\equiv t_{\nu}/(M-2k)$ appearing here can be regarded as a joint
probability over  the two variables $b$ and $j$, and we can consider  its
marginal probability $p_j\equiv p_{0j}+p_{1j}$ and the conditional
probability $p_{b|j}\equiv p_{bj}/p_j$. We use similar notations  for
other joint probabilities $q_{bb'jj'}\equiv s_{\nu\nu'}/k$, 
$\alpha_{bj}\equiv T_{\nu}/2$, and 
$\beta_{bb'jj'}\equiv S_{\nu\nu'}/4$. We further introduce
a variable $a$,
which takes three values $\{1,2,3\}$, define a probability
$\xi_a$ by $\xi_1=1-2k/M$ and $\xi_2=\xi_3=k/M$,
and define a joint probability
$\gamma_{ab}$ over $a$ and $b$, defined by $\gamma_{1b}=\xi_1p_b$,
$\gamma_{2b}=\xi_2q_b$,  and $\gamma_{3b'}=\xi_3q_{b'}$. Then, it is a
bit tedious but straightforward to 
 rewrite Eq.~(\ref{R1}) as
\begin{eqnarray} &&R=(k/M)[D(q_{bb'}|q_bq_{b'})+\sum_{bb'}
q_{bb'}D(q_{jj'|bb'}|\beta_{jj'|bb'})]
\nonumber \\ &&+(1-2k/M)\sum_{b}
p_{b}D(p_{j|b}|\alpha_{j|b})+D(\gamma_{ab}|\gamma_a\gamma_{b})
\end{eqnarray} where we have used $\gamma_b=M_b/M$, $\alpha_b=1/2$ and
$\beta_{bb'}=1/4$. Since all terms are nonnegative, $R$ is zero only if
each pair of  probabilities in $D$ are identical. This implies
$p_{bj}= |\langle b,j |{\bf n}\rangle|^2(M_b/M)$,
$q_{bj}=(1/2)(M_{b}/M)$, and $q_{b'j'}=(1/2)(M_{b'}/M)$. 
From the relation $n_{b,j}=M(\xi_1p_{bj}+\xi_2q_{bj}
+\xi_3q_{b'j'}|_{b'=b,j'=j})$
we conclude that,
for ${\rm min}\; R$ to be zero, it is
necessary that 
\begin{equation}
\delta_b=\xi_1|\langle b,1 |{\bf n}\rangle|^2 + (1-\xi_1)/2
\end{equation}
 for a choice of $|{\bf n}\rangle$ and $0\le \xi_1\le 1$,
or equivalently, $\delta_b=\langle b,1 |\sigma|b,1\rangle$
for a state $\sigma$ of a single qubit.
Otherwise, $p(\delta_0,\delta_1)$ is as
exponentially small as $\exp[- M \min R]$. Note that 
in the limit of 
$M\rightarrow\infty$, the result is consistent with 
what is expected from the quantum de Finetti theorem \cite{cfs01}.

 Now applying this general result to our case, we have
\begin{equation} 
\sin^2(\theta_l-\theta)-\epsilon_7\le \sin^2\phi_l \le
\sin^2(\theta_l+\theta)+\epsilon_8
\label{lastineq}
\end{equation} 
for $l=0,1$, where all the angles are defined in $[0,\pi/2]$ 
by the relations
$n_{11}/(n_{11}+n_{00})=\sin^2\theta_0$,
$n_{01}/(n_{01}+n_{10})=\sin^2\theta_1$, 
$m_{11}/(m_{11}+m_{00})=\sin^2\phi_0$,
$m_{01}/(m_{01}+m_{10})=\sin^2\phi_1$, 
and
$\alpha^2=\sin^2\theta$. Together with 
Eqs.~(\ref{ineq2})--(\ref{ineq6}), 
an exponentially-reliable upper bound
of $n_{\rm ph}$ can be found.

\begin{figure}[tbp]
\begin{center}
 \includegraphics[scale=0.5]{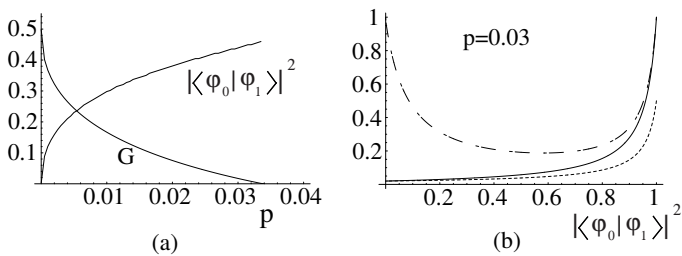}
\end{center}
 \caption{\setlength{\baselineskip}{1.0mm} (a) The optimum value of $|\langle{\varphi_0}|{\varphi_1}\rangle|^2$ 
and the key generation rate $G$ in the depolarizing
channel. (b) The error rates (normalized by $n_{\rm fil}$)
in the data qubits 
for the depolarizing channel with $p=0.03$. 
The estimated upper bound for phase errors (dot-dashed),
the actual phase errors (solid), and 
the bit errors (dotted).
\label{keygain}}
\end{figure}

In the following, we calculate
 the final key length in the limit of large
$N$, by setting all $\epsilon_j$ to be zero.
From Eq.~(\ref{ineq1}), $n_{\rm bit}$ is found to be 
equal to $n_{\rm err}$.
  Eqs.~(\ref{ineq2})--(\ref{ineq6}) are now linear 
equations, and together with the relation
$\sum n_{ij}=\sum m_{ij}=N$, they
 can be used to eliminate $n_{ij}$ and
$m_{ij}$. Then, the inequalities (\ref{lastineq})
for $l=0,1$ are combined to give
\begin{equation}
|n_{\rm fil}-2n_{\rm err}|\le N\alpha\beta f(x),
\label{ineqlimit}
\end{equation}
where 
$f(x)\equiv\sqrt{x^2-\Delta^2}+\sqrt{(1-x)^2-(\beta^2-\alpha^2-\Delta)^2}$
with
$\Delta\equiv (n_{\rm fil}/N-2\alpha^2\beta^2)/(\beta^2-\alpha^2)$ and
$x\equiv 2 n_{\rm ph}/N-(\beta^2-\alpha^2)\Delta$. The positivity of
$n_{ij}$ requires that 
$|\Delta|\le x \le 1-|\beta^2-\alpha^2-\Delta|$.
Solving Eq.~(\ref{ineqlimit}) gives an upper bound 
$\overline{n}_{\rm ph}$ of the number of phase errors  
$n_{\rm ph}$, as a function of 
the observed values $n_{\rm err}$ and $n_{\rm fil}$.

The achievable length of the final key is given \cite{CSS, GLLP02}
by 
$n_{\rm key}=n_{\rm{fil}}[1-h(n_{\rm bit}/n_{\rm fil})-h(\overline{n}_{\rm
ph}/ n_{\rm fil})]$, when $\overline{n}_{\rm ph}/
n_{\rm fil}\le 1/2$
 [note that positions of errors are
randomized in step (2)]. Here $h(p)\equiv H(p,1-p)$.
 In order to show a quantitative example of the security, we assume 
 that the channel
is the depolarizing channel where the state $\rho$ evolves as
$\rho\rightarrow (1-p)\rho +p/3\sum_{a=x,y,z}\sigma_{a}\rho\sigma_{a}$,
where $\sigma_{a}$ is the Pauli operator of $a$ component.
In Fig.~\ref{keygain}(a), we plot the key generation rate 
$G=n_{\rm key}/N$ optimized over the nonorthogonality $|\langle{\varphi_0}|{\varphi_1}\rangle|^2$.
 It is seen that our protocol is secure up to $p\sim0.034$, which is
 smaller than in BB84 with one-way EDP 
($p\sim0.165$) \cite{sp00}. In Fig.~\ref{keygain}(b), it can be seen that
when $|\langle{\varphi_0}|{\varphi_1}\rangle|^2$ becomes smaller, the estimation of the phase errors
becomes poorer. On the other hand, larger values of $|\langle{\varphi_0}|{\varphi_1}\rangle|^2$ make the signal more vulnerable to the noises, resulting 
in larger errors. This trade-off is in contrast to BB84, in which a good estimation and small errors are achieved at the same time by adding two more states in the protocol.

In summary, the B92 protocol can be regarded as an EDP with a filtering
process, and the filtering makes the phase and bit errors related to each
other, which enables us to estimate the phase errors from the amount of 
the bit errors. The estimation scheme involving nonorthogonal
measurements developed here will also be useful in practical QKD schemes
having lower symmetries due to imperfections in the apparatus.

We thank Hoi-Kwong Lo, John Preskill, and Takashi Yamamoto for helpful
discussions.

\end{document}